\documentclass[sn-mathphys,Numbered]{sn-jnl}% Math and Physical Sciences Reference Style
\usepackage[T2A]{fontenc}                        % Поддержка русских букв

\usepackage{graphicx}%
\usepackage{multirow}%
\usepackage{amsmath,amssymb,amsfonts}%
\usepackage{amsthm}%
\usepackage{mathrsfs}%
\usepackage[title]{appendix}%
\usepackage{xcolor}%
\usepackage{textcomp}%
\usepackage{manyfoot}%
\usepackage{booktabs}%
\usepackage{algorithm}%
\usepackage{algorithmicx}%
\usepackage{algpseudocode}%
\usepackage{listings}%
\theoremstyle{thmstyleone}%
\theoremstyle{thmstyletwo}%

\theoremstyle{thmstylethree}%

\newcommand{\bb}{\begin{equation}}
\newcommand{\ee}{\end{equation}}
\newcommand{\ep}{\varepsilon\,}
\newcommand{\bm}{\boldsymbol}

\begin{document}

\title{Induced electromagnetic radiation from a charged cloud in a plane gravitational wave}

%\title{Electromagnetic radiation from a plasma cloud induced by a plane gravitational waveЪ


\author[1]{\fnm{Vladimir} \sur{Epp}}\email{epp@tspu.edu.ru}

%\author*{\fnm{Konstantin} \sur{Osetrin}}\email{osetrin@tspu.edu.ru}

%\author{\fnm{Denis} \sur{Kartashov}}\email{iiiauthor@gmail.com}

\author*[1,2]{\fnm{Konstantin} \sur{Osetrin}}\email{osetrin@tspu.edu.ru}

\affil*[1]{%\orgdiv{}, 
\orgname{Tomsk State Pedagogical Unversity}, \orgaddress{\street{Kievskaya, 60}, \city{Tomsk}, \postcode{634061}, \country{Russia}}}

%\affil*[1]{
%\orgdiv{Center for Mathematical and Computer Physics}, 
%\orgname{Tomsk State Pedagogical University}, 
%\orgaddress{\street{Kievskaya str. 60}, 
%\city{Tomsk}, \postcode{634061}, 
%%\state{}, 
%\country{Russia}}}
%
%\affil[2]{
%%\orgdiv{Department}, 
%\orgname{National Research Tomsk State University}, \orgaddress{\street{Lenina pr. 36}, \city{Tomsk}, \postcode{634050}, 
%%\state{State}, 
%\country{Russia}}}

\affil[2]{
%\orgdiv{International Laboratory for Theoretical Cosmology}, 
\orgname{Tomsk State University of Control Systems and Radioelectronics}, \orgaddress{\street{Lenina pr. 40}, \city{Tomsk}, \postcode{634050}, 
%\state{State}, 
\country{Russia}}}

%\affil[2]{\orgdiv{Department}, \orgname{Organization}, \orgaddress{\street{Street}, \city{City}, \postcode{10587}, \state{State}, \country{Country}}}
%
%\affil[3]{\orgdiv{Department}, \orgname{Organization}, \orgaddress{\street{Street}, \city{City}, \postcode{610101}, \state{State}, \country{Country}}}


\abstract{
For the perturbative model of a plane gravitational wave on a flat background of Minkowski space-time, electromagnetic radiation from a charged cloud in the field of a gravitational wave, detected by a remote observer, was found. It is shown that the charge density in the cloud does not change, and the radiation is generated by currents induced by the gravitational wave. The angular distribution of the radiation is obtained. If the refractive index of the cloud medium is greater than unity, Cherenkov-type radiation is generated.
}





\maketitle

\section{Introduction}

Currently, significant research efforts are being made to develop methods for direct detection of gravitational waves. These efforts culminated in the successful registration of gravitational waves and the beginning of a new scientific field of research -- gravitational wave astronomy \cite{Abbott_2016, Abbott_2019, Abbott_2021}. However, direct detection of long-length gravitational waves faces certain difficulties. 
This is especially true for primordial gravitational waves, which can have long wavelengths and also correspond to the equations of modified theories of gravity 
\cite{https://doi.org/10.1002/prop.202100167,%
ODINTSOV2022100950,ODINTSOV2022136817,%
CAPOZZIELLO2021100867,NOJIRI2020100514,PhysRevD.98.024002,Osetrin356Universe_2023}.
Therefore, study of the alternative models for detecting gravitational waves that make it possible to observe gravitational waves by indirect methods is relevant.
In particular, a number of works are devoted to the interaction of gravitational waves with electromagnetic fields and with charged particles. The basic idea is that gravitational waves can accelerate charged particles, which should result in electromagnetic radiation. The methods for calculating this radiation and, accordingly, the calculation results are very different. In one of the first works devoted to this topic \cite{Heintzmann_1968}, a method of successive approximations for solving Maxwell's equations for a point charge in the field of a spherical  gravitational wave was proposed. A similar method is used to evaluate the interaction of a charged particle with a plane gravitational wave \cite{Wickramasinghe_2015}. It has been shown that charged particles can transform the energy of a gravitational wave into electromagnetic radiation.
Boughn \cite{Boughn_1975} has solved the Maxwell equations for a point charge  in the metric of a plane gravitational wave, expanding the electromagnetic field potential into spherical harmonics. The coefficients of this expansion are analysed using numerical methods. It is shown that the total radiation intensity summed over harmonics diverges. Methods for eliminating this divergence are proposed. The emission of a relativistic point charge colliding with a plane gravitational wave was studied by Sasaki and Sato \cite{Sasaki_1978} also using the method of successive approximations. It is shown that a charge radiates into a narrow cone in the direction of its motion. And again, in the direction of propagation of the gravitational wave, the intensity of the charge radiation diverges.
As we can see, most authors encounter difficulties associated with divergences of various kinds when calculating the electromagnetic field of a point charge in the metric of plane gravitational wave .
A number of papers are related to the construction of integrable exact models of gravitational waves, including models of primordial gravitational waves
\cite{Osetrin2022894,Osetrin325205JPA_2023,Osetrin1455Sym_2023,%
ObukhovUniverse8040245,ObukhovSym14122595,ObukhovSym15030648}

In this paper, we study the electromagnetic radiation of a continuously distributed charge in a cloud of dust or plasma, which is affected by a plane gravitational wave. The gravitational wave deforms the cloud in a known way. It is shown that the charge density in the cloud does not change, but the deformation of the cloud induces periodically changing currents that generate electromagnetic radiation. The properties of this radiation were studied depending on the dielectric constant of the  cloud medium. 
It is shown that, under certain conditions, a gravitational wave can induce Cherenkov radiation from a plasma cloud.
%%%%%%%%%%%%%%%%%%%%%%%%%%%%%%%%%%%%

\section{A cloud of charged particles in the gravitational wave}

Let a plane gravitational wave incident on a cloud  with  uniformly distributed charge. We take  metric the gravitational wave in the form
\begin{align}
g^{\mu\nu}&=\eta^{\mu\nu}+h^{\mu\nu},\quad h\ll 1,
\label{g}
\\
h^{\mu\nu}&=a^{\mu\nu}\exp(i\kappa_\sigma x^\sigma),
\end{align}
where $\eta^{\mu\nu}$ is the Minkowsky space metric and $a^{\mu\nu}$ is the wave amplitude. In the transverse-traceless gauge   $a^{\mu\nu}$ can be written as
\bb\label{matr}
 a^{\mu\nu}=
 \begin{bmatrix}
    0       &0 &0 & 0 \\
   0  & a & b &0\\
    0&b&-a&0 \\
   0& 0 & 0 &0
\end{bmatrix}
 \ee
for a wave travelling in the $x^3$-direction. 

Let us consider the deformation of a charged cloud under the influence of a gravitational wave. We will assume that the cloud is of cylindrical shape and the axis of the cylinder coincides with the direction of wave propagation.
The time dependance of the particles positions we find as solution to the geodesic equation 
with 4-velocity~$u^\nu$ 
\bb
\frac{du^\nu }{d\tau}+\Gamma ^\nu_{\mu\sigma}u^\mu u^\sigma=0.
\ee
In the same coordinates as used in (\ref{matr}), let them be $\xi^\nu$,  the solution to geodesic equation reads $u^\nu=\dot\xi^\nu= (c,0,0,0)$. Hence, the space 3-vector $\bm\xi^i$ ($i =1,2,3$) is constant.
Next, we introduce the spatial coordinates $\zeta^i$, which represent the physical separation of nearby particles in the gravitational wave  (see for example \cite{Hobson_2006})
\bb
\zeta^i=\xi^i+\frac 12 h^i_k\xi^k, \quad i,k =1,2,3
\ee
For simplicity, we will assume that the gravitational wave is polarized so that in the tensor (\ref{matr}) $b=0$.
Substituting $h^i_k$ into the last equation, we find that in the plane orthogonal to the wave vector of the gravitational wave, the particle coordinates, expressed in units of length, vary as \cite{Hobson_2006}
\begin{align}
\zeta^1&=\xi^1
\left[
1-\frac{a}{2}\cos{\kappa(ct-\xi^3)}
\right]
\label{11}\\
\zeta^2&=\xi^2
\left[
1+\frac{a}{2}\cos{\kappa(ct-\xi^3)}
\right]\\
\zeta^3&=\xi^3
\label{21}
\end{align}
The constant vector $\bm\xi=(\xi^1,\xi^2,\xi^3)$ labels the initial position of the particle. 
The vector field of particle velocities has the form
\bb
\bm v=
\frac{a}{2}\,\omega \left(\xi^1,-\xi^2,0\right)\sin \kappa(ct-\xi^3)
\label{31}
.\ee
Accordingly, the electric current density in the cloud is equal to $\bm j =\rho \bm v$. Variation of the  charge density in the cloud we find from the continuity equation \cite{Landau_II}
\bb
(-g)^{-1/2}\partial_\alpha( \sqrt {-g}\,j^\alpha)=0
\label{42}
,\ee
where $g$ is the determinant of the metric tensor. Eq. (\ref{matr}) shows that $g$ differs from minus one by terms of second order of smallness in $h$. Hence, up to  the first order  in $h$, the continuity equation reads
\bb
\mbox{div} (\rho\bm v)+\frac{\partial\rho}{\partial t}=0
\label{41}
.\ee
We look for a solution for $\rho$ in the form of a Fourier series expansion in time. As a result, we get $\rho=const$.

Thus, the effect of a gravitational wave on a cloud is that the charges are displaced in a plane orthogonal to the direction of propagation of the wave, the charge density does not change, but the current distributed in the cross section of the cloud  is induced according to the law (\ref{31}). This alternating current can be expected to generate electromagnetic radiation. 

If we restrict ourselves to the first order approximation in $h$, then the further calculation of the electromagnetic field produced by currents and the propagation of electromagnetic waves can be carried out as for the flat space with the metric $\eta_{\mu\nu}$.
This can be seen, for example, from the following reasoning. Maxwell's equations in a gravitational field can be written as equations in a material medium with a certain dielectric and magnetic permeabilities \cite{Landau_II}.
These characteristics of the medium are expressed through the determinant of the metric tensor. However, as we have already noted, the determinant of the metric tensor (\ref{g}) is equal to minus one up to terms of order $h^2$. Therefore, in the linear approximation in $h$, Maxwell's equations in the metric (\ref{g}) coincide with Maxwell's equations in flat space.
%%%%%%%%%%%%%%%%%%%%%%%%%%
\section{Radiation from the charged cloud}
%%%%%%%%%%%%%%%%%%%%%%%%%

At large distances from the region where the radiation is generated, the electric field of the radiation is set only by the vector potential
$\bm A(\bm r,t)$ \cite{Landau_II}
\bb\label{E}
\bm E=((\bm{\dot A}\times\bm e)\times\bm e)
\ee 
where $\bm e$ is the unit vector in the direction of radiation, the dot denotes the time derivative $t$. The vector potential is determined by the current density at a delayed moment in time
\bb
\bm A=\frac{1}{cR_0}\int{\bm j \left(t-\frac{R_0}{c}+\frac{\bm r \bm e\sqrt{\ep}}{c}
\right)
}\,dV
\label{51}
.\ee
Here $R_0$ is the distance from the center of the cloud to the observer, $\sqrt{\ep}$ is the dielectric constant of the medium in the cloud, $\bm r$ is the radius vector of the volume element $dV$. We wrote the fraction $R_0/c$ without dielectric constant, because we believe that $R_0$ is much larger than the size of the cloud and the radiation propagates from the cloud to the observer in a vacuum.

Let us represent the vector potential in the spherical coordinate system $(R_0,\theta,\phi)$ as $\bm A= (A_R,A_\theta, A_\phi)$. Obviously, $A_R$ is not included in the formula (\ref{E}), so it is enough to calculate $A_\theta$ and $A_\phi$. The current density components in the spherical coordinate system are equal
\begin{align}
j_\theta&=
j_1\cos{\theta}\cos{\phi}+j_2\cos{\theta}\sin{\phi}\\
j_\phi&=-j_1\sin{\phi}+j_2\cos{\phi}
\label{61}
.\end{align}

Substituting the current density $\bm j=\rho\bm v$ into equation (\ref{51}) in accordance with equation (\ref{31}) and integrating over the volume, we obtain
\begin{align}
A_\theta&=
\frac{2\pi a {r_0}^2 \cos{\theta}
}{
R_0\omega n (1-n\cos{\theta})\sin{\theta}
}\,
\sin{\left[
\frac{L\omega}{2c}(1-n\cos{\theta})
\right]}e^{i\omega t}
J_2(z),
\label{71}\\
A_\phi&=0,\nonumber\\
z&=\frac{\omega n r_0 }{c}\sin\theta
.\end{align}
Here $J_2(z)$ is the Bessel function, $n=\sqrt{\ep}$ is the refractive index, $r_0$ and $L$ are the radius and the length of the cloud respectively. We see that the frequency of radiation is the same as the frequency of gravitation wave. 

When integrating over volume in equation (\ref{51}), we have neglected the variation of the cloud surface with time. The amplitude of this variation is proportional to $h$, however, the current density $j(t)$ is already proportional to $h$. So the pulsation of the cloud surface adds a second-order correction to the integral.

It follows from (\ref{E}) that $E_\theta=-A_\theta, \, E_\phi=0$. Finally, we calculate intensity of radiation as  time-dependent intensity averaged over period
\bb
\frac{d I}{d\Omega}=\frac{c}{4\pi}\bar{E^2}R_0^2=
\frac{
\pi c a^2 r_0^4
\sin^2{\left[
\frac{L\omega}{2c}(1-n\cos{\theta})
\right]}
}{
2n^2 \sin^2{\theta}\,(1-n\cos{\theta})^2
}
\, J_2^2(z)
\label{111}
.\ee

To facilitate the analysis of the obtained result, we denote by $\eta_\bot$ the dimensionless radius of the cloud in units of the gravitational wave length and by $\eta_\parallel$ the length of the cloud in the same units. The cloud volume $V$ can also be expressed in dimensionless units.
\bb
\eta_\bot=\frac{r_0\omega}{2\pi c},\quad \eta_\parallel=\frac{L\omega}{2\pi c}, \quad V=\pi \eta_\bot^2\eta_\parallel
\ee
Then the radiation intensity will take the form
\bb
\frac{d I}{d\Omega}=\frac{
8\pi^3 c^5 a^2 V^2}{
n^2\omega^4 \sin^2{\theta}}\,
\frac{\sin^2{\left[\pi\eta_\parallel
(1-n\cos{\theta})
\right]}}{\eta_\parallel^2(1-n\cos{\theta})^2}
\, J_2^2(2\pi\eta_\bot n\sin\theta)
\label{11-1}
.\ee
The factor $V^2$ in the numerator of this expression indicates that the radiation in the cloud is generated coherently. This is a consequence of the fact that the gravitational wave excites currents in the cloud in a consistent manner. Another consequence of this fact is the modulation of the angular distribution along the angle $\theta$, which is reflected by the square of the sine in the numerator. In other words, the radiation pattern is, generally speaking, multi-lobed. There is no radiation in the direction of propagation of the gravitational wave ($\theta=0$), since at $\theta\to 0$ the square of the Bessel function tends to zero as $\theta^4$.

As can be seen from equation (\ref{71}), the  radiation is polarized in a plane passing through the cloud axis and the point where the observer is located. This is a consequence of the fact that we considered a polarized gravitational wave. Naturally, in the general case of a monochromatic gravitational wave, electromagnetic radiation will have elliptical polarization.

Note, that the angular distribution of intensity  contains a factor $1-n\cos{\theta}$ in the denominator.
If the index of refraction is greater than unity ($n>1$), then the radiation is of  Cherenkov radiation type. This could be expected because the gravitational wave propagates in the cloud faster than the speed of light in this medium. 
However, unlike the Cherenkov radiation, expression (\ref{11-1}) remains finite at $1-n\cos{\theta}=0$. And only at $L\to\infty$, as can be seen from equation (\ref{111}), the angular distribution of radiation degenerates into a delta function.
Radiation in this case is confined within the Cherenkov cone with the opening angle $\theta_c $, such that $\cos{\theta_c}=n^{-1}$ and propagates outward  at the angle $\theta_c $.

\section{Discussion}

We considered a very simple model of the interaction of a gravitational wave with charged matter in order to identify the basic properties of the resulting electromagnetic radiation during the collective motion of charged particles. 
In the papers cited in the introduction, only radiation of individual charged particles was studied.

Since we were mainly interested in the mechanism of electromagnetic radiation and the properties of this radiation, we do not discuss here the issue of how widespread clouds of plasma, dust or gas with an uncompensated electric charge are in interstellar space. We only note that extensive regions with separated charge can appear at the front of a shock wave during a supernova explosion, in relativistic jets and in the vicinity of neutron stars if the star’s rotation axis does not coincide with the magnetic axis. Also the dust clouds cannot be neglected because dust particles and electrons move in the gravitational field with the same acceleration, but he grains of dusty clouds can carry a significant electrical charge \cite{Bingham_2001, Tsytovich_2014}.

In a highly rarefied cloud, when the distance between charged particles is greater than the radiation wavelength, the particles emit incoherently. In this case, even a cloud that is neutral on average will generate radiation \cite{Boughn_1975}.

The results obtained here are obviously not applicable for opaque clouds when the frequency of the electromagnetic wave is less than the Langmuir frequency of plasma oscillations.

\section{Conclusion}
A perturbative model of radiation from a charged cloud in a plane gravitational wave is considered. Electromagnetic radiation from a charged cloud in the field of a gravitational wave, recorded by a remote observer, was found. It is shown that the charge density in the cloud does not change, and the radiation is generated by currents induced by the gravitational wave. The angular distribution of radiation has been found. It is shown that if the refractive index of the cloud medium is greater than unity, Cherenkov-type radiation is generated.

\section*{Acknowledgments}
The study was supported by the Russian Science Foundation, grant \mbox{No. 23-22-00343},
\\
\url{https://rscf.ru/en/project/23-22-00343/}.

%\bibliographystyle{unsrt}
%\bibliography{eppbibfile_gr}
%\bibliography{eppbibfile_gr_2023}


\end{document}